\documentclass[%
  reprint,
  aps,prb,
  amsmath,amssymb,
  superscriptaddress,
  floatfix
]{revtex4-2}

\usepackage[T1]{fontenc}
\usepackage[utf8]{inputenc}

\usepackage{newtxtext}
\usepackage{newtxmath}

\usepackage{microtype}

\usepackage{graphicx}
\usepackage{dcolumn}
\usepackage{bm}
\usepackage[version=4]{mhchem}

\usepackage[hidelinks]{hyperref}

\begin{document}

\preprint{APS/123-QED}

\title{Controlling the Spin-Wave Nonreciprocity of a Crescent-Shaped Nanowire via Curvature and Magnetic Field}

\author{Uladzislau Makartsou}
\email{ulamak@amu.edu.pl}
\affiliation{Institute of Spintronics and Quantum Information, Faculty of Physics and Astronomy, Adam Mickiewicz University, Uniwersytetu Poznańskiego 2, 61-614 Poznań, Poland}
\author{Mateusz Gołębiewski}
\affiliation{Institute of Spintronics and Quantum Information, Faculty of Physics and Astronomy, Adam Mickiewicz University, Uniwersytetu Poznańskiego 2, 61-614 Poznań, Poland}
\author{Attila K\'akay}
\affiliation{Helmholtz--Zentrum Dresden--Rossendorf, Institute of Ion Beam Physics and Materials Research, Bautzner Landstraße 400, 01328 Dresden, Germany}
\author{Olena Tartakivska}
\affiliation{Institute of Spintronics and Quantum Information, Faculty of Physics and Astronomy, Adam Mickiewicz University, Uniwersytetu Poznańskiego 2, 61-614 Poznań, Poland}
\affiliation{V. G. Baryakhtar Institute of Magnetism of the NAS of Ukraine, 36b Vernadsky Boulevard, 03142 Kyiv, Ukraine}
\author{Maciej Krawczyk}
\affiliation{Institute of Spintronics and Quantum Information, Faculty of Physics and Astronomy, Adam Mickiewicz University, Uniwersytetu Poznańskiego 2, 61-614 Poznań, Poland}

\date{\today}

\begin{abstract}
Recent studies on spin-wave propagation in ferromagnetic waveguides has highlighted the role of nonreciprocity resulting from the chiral nature of dipolar interactions in curved elements. However, the impact of spin-wave mode type on nonreciprocity remains unexplored. Using micromagnetic simulations supported by analytical modeling, we systematically analyzed the propagation of edge, fundamental, and width-quantized spin-wave modes in a ferromagnetic nanowire with a crescent-shaped cross-section. Our results show that the strength and sign of nonreciprocity depend on the mode type, as well as on the curvature magnitude of the nanowire's top and bottom surfaces and the strength of the external magnetic field. Interestingly, changing the mode type, for instance induced by altering the curvature or magnetic field, result in a significant change in the dispersion relation asymmetry. This effect underscores the important role of spin-wave profiles in nonreciprocity, deepens our fundamental understanding of spin-wave dynamics in curved geometries, and paves the way for designing magnonic waveguides with tailored properties.

\end{abstract}

\maketitle

\section{Introduction\label{Sec:Introduction}}
Progress in nanoscale fabrication technologies improves the ability to control and manipulate spin waves (SW)~\cite{Gubbiotti_2019, Fischer_2019}. Such control can be achieved through geometrical factors---including the size, shape, and curvature of the ferromagnet---as well as through the orientation and texture of the magnetization~\cite{Donnelly2021ComplexNanostructures, Girardi2024}. For instance, self-assembly, or two-photon lithography for polymer structuring, combined with electrodeposition~\cite{Golebiewski2024,Hunt2020HarnessingNanoscale, vandenBerg2022CombiningNanostructures}, atomic layer deposition (ALD)~\cite{Guo_2023}, or focused electron beam induced deposition (FEBID)~\cite{Fernandez_2020}, 
enables the realization of nearly arbitrary shapes in the sub-micron regime~\cite{Makarov2022NewNanoarchitectures, Fernandez-Pacheco2017Three-dimensionalNanomagnetism}. Understanding how the geometric and topological properties of ferromagnetic materials influence SW propagation in sub-micron systems is therefore emerging as a key frontier in magnonics~\cite{Gubbiotti_2025}.

Even a single nanowire, acting as an SW waveguide, provides a versatile platform for studying complex magnetization dynamics and exploring potential applications~\cite{STANO2018, Biorn2020, Giordano_2023}. When curvature is introduced, either along the wire axis \cite{Tkachenko2012} or across its cross-section, additional phenomena emerge~\cite{Streubel2016MagnetismGeometries}, including domain-wall pinning or curvature-induced interactions reminiscent of the Dzyaloshinskii--Moriya type, respectively~\cite{Sheka2021AMagnetism}. Among these, nonreciprocity is of particular interest. As a manifestation of broken time-reversal symmetry, nonreciprocity of SWs has been recognized for decades~\cite{CAMLEY1987103}, and the ability to control it is crucial for device functionalities such as isolators, diodes, and circulators~\cite{PhysRevApplied.14.034063,Jamali2013}, which are essential in wave-based signal processing. However, studies of SW dynamics in this context have so far been limited to thin films and bilayers~\cite{PhysRevB.49.339, Szulc2024,10.1063/10.0038642,Ishibashi2020, PhysRevB.111.134434,l93m-gb54}, and nanowires with highly symmetric cross-sections and regular magnetization textures, such as tubes with circular or hexagonal cross-section, or rolled circular tubes~\cite{Otalora2016, korberSymmetryCurvatureEffects2021}. Moreover, these studies are focused only on the fundamental SW mode. 
 
Nanowires with a crescent-shaped (CS) cross-section (see Fig.~\ref{Fig:structure}), including those arranged in diamond-like networks~\cite{May2019realizationLattice,  May2021MagneticSpin-ice, Sahoo2018UltrafastStructure}, have already proven promising for realizing 3D artificial spin-ice structures and for providing reconfigurable platforms for magnonic devices~\cite{Sahoo2021ObservationStructure,PhysRevApplied.19.064045}. Following these works, standalone CS nanowires were also realized with two-photon lithography combined with thermal evaporation of Py~\cite{Askey2024}. Subsequently, straight nanowires of the similar CS shape were fabricated in the inverse arrangement, i.e., by electron-beam lithography, reactive-ion etching, and Py sputtering on pre-patterned Si substrates~\cite{Pradhan2024ICM}. 
In addition, deposition of magnetic layers onto nanospheres, establishes curved geometries in the form of ferromagnetic cups~\cite{Albrecht2005}. This technique can be extended to coat a polymer nanorod with ferromagnet and form long, CS ferromagnetic nanowires. Within this framework, CS nanorods can represent a fundamental building block for magnonic architectures. Despite this potential, the nonreciprocal properties of SWs in CS nanowires have not yet been analyzed.  

In this work, we study SW nonreciprocity in an infinitely long nanowire with a CS cross-section made of Py. The assumed geometry introduces unique effects, such as varying curvature of the top and bottom surface in the cross-section, inhomogeneous magnetization, and edge contributions that remain absent in nanowires with highly symmetric cross-sections. This allows us to expand our analysis  beyond the fundamental mode, which is defined by in-phase oscillations throughout the entire cross-section, and was partially covered in Ref.~\cite{PhysRevApplied.19.064045}. We include edge-localized and quantized modes in the transverse plane. We show that the sign and magnitude of the nonreciprocity depend not only on the nanowire curvature and magnetization state but also on the mode character, exhibiting strong variation with the wavenumber. The micromagnetic simulation results are supported by an analytical model, 
which clarifies how inversion-symmetry breaking and 
mode profile govern the nonreciprocity of SWs propagating in transversely magnetized nanowires. The analyzed modes can be excited and detected using a microstrip antenna, which provides direct relevance to an experimental realization. Therefore, the reported properties of SWs in the CS nanowire demonstrate potential for nanoscale magnonics.

\section{Geometry, methods and magnetization ground state\label{Sec:SimGeometryParams}}

\begin{figure}[ht]
\centering
\includegraphics[width=\columnwidth]{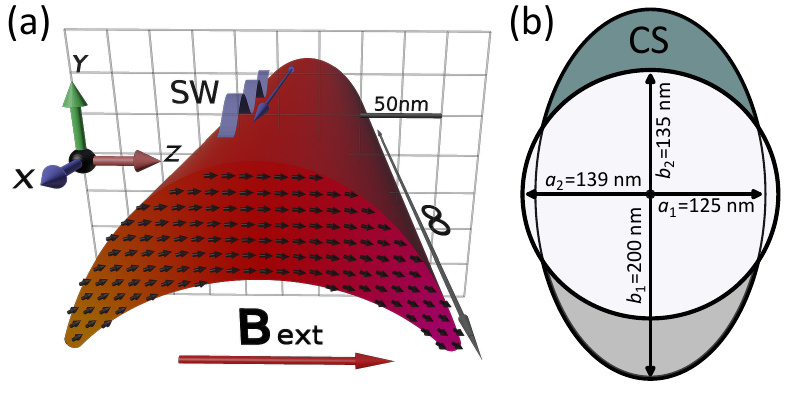}
\caption{(a) Infinitely long CS nanowire with an external magnetic field \(B_{\text{ext}}=0.5\)~T applied along the $z$-axis. Arrows indicate the steady-state magnetization distribution. (b) Geometrical construction of the CS cross-section, defined as the set difference of two ellipses: \(E_1\) with semi-axes $a_1=125$~nm and $b_1=200$~nm, \(E_2\) with semi-axes $a_2=139$~nm and $b_2=135$~nm.}
\label{Fig:structure}
\end{figure}

We employ the TetraX micromagnetic modeling package~\cite{korber_tetrax} to determine the equilibrium state and to calculate the SW dispersion in infinitely long CS nanowires. The software implements a finite-element dynamic-matrix approach in 2D (the nanowire cross-section). The equilibrium magnetization distribution is obtained by minimizing the total energy density using a conjugate-gradient method~\cite{10.1063/5.0054169}. To evaluate SW spectra, the solver addresses the linearized Landau–Lifshitz–Gilbert (LLG) equation formulated as an eigenvalue problem parameterized by the SW wavenumber \(k_x\) along the nanowire axis~\cite{Körber2022,10.1063/5.0054169}. The finite-element approach provides an accurate reconstruction of the CS edges. The mesh of the structure was generated in COMSOL Multiphysics and exported to TetraX, employing an inhomogeneous discretization refined at the edges, with a maximum element size of 1~nm, which is well below the exchange length of permalloy (Py) (\(\approx 5.7~\mathrm{nm}\)). 

The cross-sectional geometry of the CS nanowire is defined as the set difference between two intersecting ellipses, \(\text{CS} = E_1 \setminus E_2\) [see Fig.~\ref{Fig:structure}(b)], where \(E_1\) and \(E_2\) denote ellipses with semi-major axes \(b_i\) and semi-minor axes \(a_i\) (\(i = 1, 2\)). To avoid singularities and better reflect experimental realizations~\cite{May2019realizationLattice}, the edges were rounded with a 5~nm radius [Fig.~\ref{Fig:structure}(a)]. Unless otherwise stated, the geometry parameters follow those of Ref.~\cite{PhysRevApplied.19.064045}: \(a_1 = 125~\mathrm{nm}\), \(b_1 = 200~\mathrm{nm}\), \(a_2 = 139~\mathrm{nm}\), and \(b_2 = 135~\mathrm{nm}\). In addition, variation of these parameters were considered to examine the influence of shape modifications on SW nonreciprocity. Throughout this study, the nanowires are treated as infinitely long, such that magnetic properties (e.g., magnetization, demagnetizing field) are assumed to be translationally invariant along the \(x\)-axis. For the material parameters, we assume Py, characterized by values standard for Py thin films: saturation magnetization $M_{\text{s}}$ = 800~kA/m, exchange stiffness constant $A_{\text{ex}}$ = 13~pJ/m, Gilbert damping constant $\alpha = 0.008$, and reduced gyromagnetic ratio $\gamma/2\pi$ = 28~GHz/T.

\begin{figure}[h]
\centering
\includegraphics[width=\columnwidth]{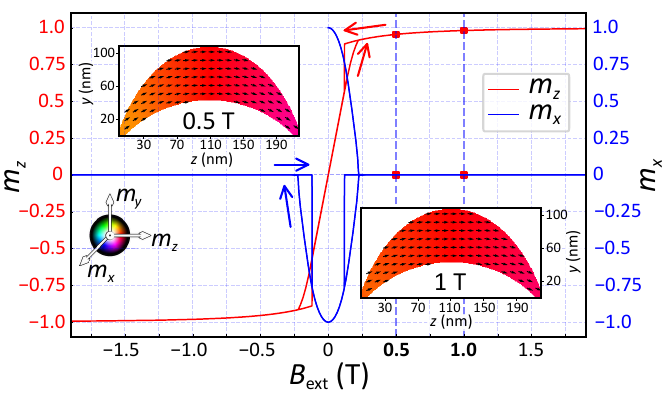}
    \caption{Hysteresis loops of the CS nanowire illustrating the normalized, cross-sectionally averaged magnetization components, $m_z$ (red line) and $m_x$ (blue line), as a function of the external magnetic field $B_\mathrm{ext}$ applied along the $z$-axis. Arrows indicate the loop orientation for increasing and decreasing field. Insets display the static magnetization configurations at $B_\mathrm{ext} = 0.5$ and 1~T, with the colormap representing magnetization orientation as defined in the diagram in the lower-left corner.}
\label{Fig:hyst}
\end{figure}

Prior to calculating the SW spectra, we examine the static magnetization configuration under an external magnetic field $B_{\mathrm{ext}}$. The simulated hysteresis loop for a field applied along the $z$-axis is shown in Fig.~\ref{Fig:hyst}, where the red line corresponds to the $z$-component of the magnetization. The curve exhibits the characteristic shape of a hard-axis orientation, with a narrow hysteretic region near the switching field ($\approx \pm 0.13$~T). In this range, the magnetization begins to reorient towards the transverse direction, as indicated by the blue curve showing $m_x(B_{\mathrm{ext}})$ (the normalized $x$-component of the magnetization). For $|B_{\mathrm{ext}}|>0.25$~T, the magnetization lies entirely within the cross-sectional plane, and this field region is chosen for the subsequent analysis of SW dynamics.

\section{Results\label{Sec:Results}}
\begin{figure*}[hbt]
\centering
\includegraphics[width=\linewidth]{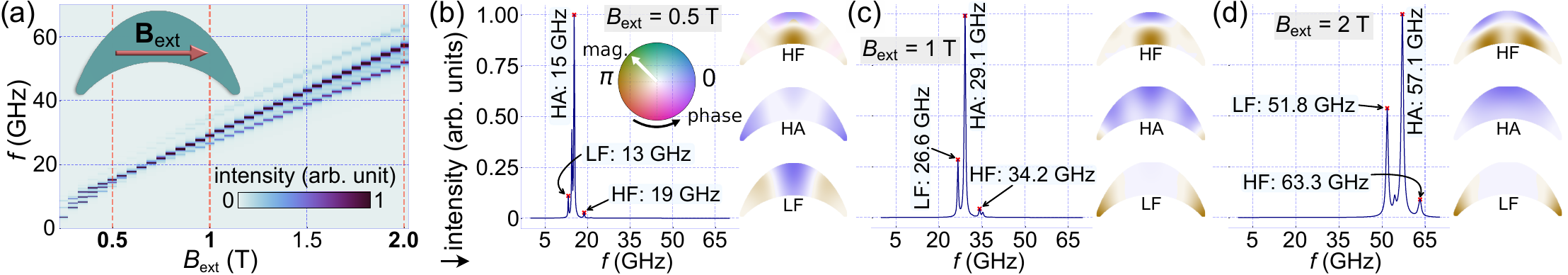}
    \caption{(a) FMR spectra of the CS nanowire as a function of the external magnetic field $B_{\mathrm{ext}}$ applied along the $z$-axis. Color represents the normalized absorption intensity (scale bar), and the red lines mark the field values taken into closer analysis in (b)-(d). (b)--(d) Spectra at selected field values: $B_{\mathrm{ext}}=$ 0.5~T, 1~T and 2~T, respectively. Insets LF, HA, and HF show the corresponding SW mode profiles: hue encodes the local precession phase (color wheel), while opacity encodes the normalized  amplitude $|m| = \sqrt{|m_x|^2 + |m_y|^2}$ (transparent at 0, opaque at 1).}
    \label{Fig:FMR}
\end{figure*}

\subsection{Ferromagnetic resonance analysis\label{Sec:FMR}}

As a first step, we compute the ferromagnetic resonance (FMR) spectra in the relevant range of external magnetic fields applied along the $z$-axis. 
The simulations involve calculating the complex magnetic susceptibility \(\chi\), where peaks in the imaginary component \(\chi''\) correspond to the resonance frequencies of maximum energy absorption, hereafter referred to as FMR. The susceptibility calculation involves all dynamic components of the magnetization. However, for $|B_{\mathrm{ext}}|>0.25$~T when magnetization lies predominantly along the $z$-axis, the contribution of $m_x$ and $m_y$ dominates. The corresponding results are presented in Fig.~\ref{Fig:FMR}(a). 
The resonance frequencies of all modes increase with the applied magnetic field. For $B_\text{ext} \geq 0.5$~T, individual modes can be clearly tracked with increasing field strength, whereas below this threshold the magnetization undergoes rotation in the $(x,y)$ plane, leading to mode mixing. For detailed analysis, we focus on three representative modes with pronounced intensity and distinct character: a low-frequency (LF) mode, a high-amplitude (HA) mode, and a high-frequency (HF) mode. These are shown in Fig.~\ref{Fig:FMR}(b--d) for fields of 0.5, 1.0 and 2.0~T, together with corresponding FMR spectra. Each of these modes is symmetric with respect to the central line of the crescent cross-section parallel to the $y$-axis and:
\begin{itemize}
\item 
The HA mode exhibits a uniform phase across the cross-section and can therefore be identified as the fundamental mode. Its SW amplitude distribution, however, evolves with the bias magnetic field strength~\cite{PhysRevApplied.19.064045}: at high fields (e.g., $B_{\mathrm{ext}}$ = 2~T) it is centered in the bulk [Fig.~\ref{Fig:FMR}(d)], while at lower fields (e.g., $B_{\mathrm{ext}}$ = 0.5~T) the maximum amplitude shifts towards the edges [Fig.~\ref{Fig:FMR}(b)]. 
\item
In contrast, the spatial distribution of the LF mode evolves from an edge-localized state, with maximum amplitude at the crescent boundaries at high fields, to a mode concentrated in the central region of the cross-section at low fields [see Fig.~\ref{Fig:FMR}(b--d)]. In this configuration, the central region  oscillates in opposite phase with respect to the edges, resulting in two nodal lines oriented parallel to the $y$-axis, as shown in Fig.~\ref{Fig:FMR}(b].
\item
The HF mode is characterized by a nodal line parallel to the $z$-axis and remains qualitatively unchanged under variations of the external magnetic field $B_{\mathrm{ext}}$. 
\end{itemize}
We attribute the changes in the HA and LF mode profiles observed at a decreasing bias field to the rotation of the static magnetization at the edges of the CS, which is forced by the demagnetizing field, as will be discussed yet in the following section.
 
\subsection{Nonreciprocity}

\subsubsection{Dependence on the bias magnetic field}

\begin{figure*}[t]
\centering
\includegraphics[width=\linewidth]{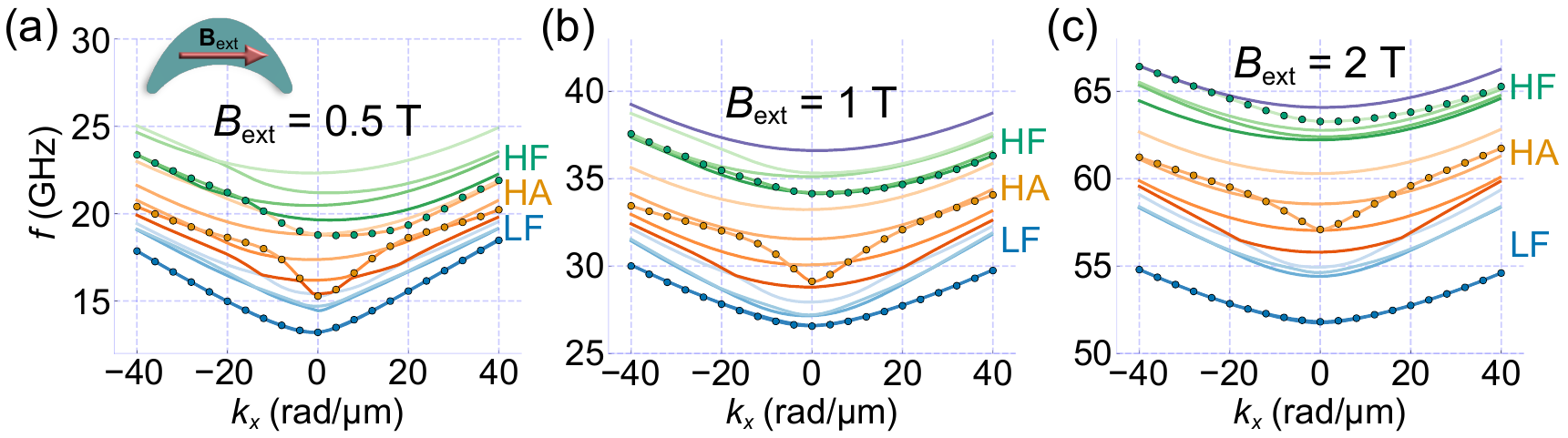}
\caption{\label{Fig:dispersions}
Dispersion relations of the first, counting from the lowest frequency, 16 SW modes propagating along the CS nanowire 
at three different values of the external magnetic field applied perpendicular to the nanowire axis. 
The LF, HA, and HF modes identified in Fig.~\ref{Fig:FMR} are highlighted with blue, orange, and green dots, respectively.}
\end{figure*}

\begin{figure}[ht]
\centering
\includegraphics[width=\columnwidth]{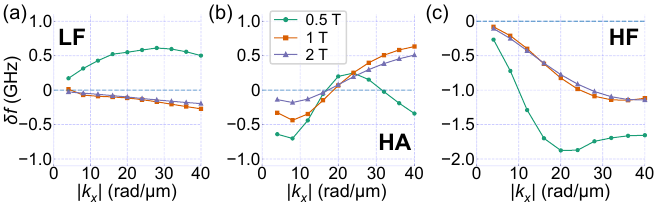}
\caption{Nonreciprocity $\delta f(|k_x|) = f(-|k_x|) - f(|k_x|)$, i.e., the frequency difference between opposite wave vectors as functions of wavenumber $|k_x|$, for the three selected in Fig.~\ref{Fig:FMR} SW modes: LF (a), HA (b), and HF (c). Results are shown for three values of the bias magnetic field applied perpendicular to the nanowire axis with magnitude $B_{\mathrm{ext}}=$~0.5, 1.0 and 2.0~T, at which dispersion relations are shown in Fig.~\ref{Fig:dispersions}.}
\label{Fig:NRP_modes_diff}
\end{figure}     

To investigate nonreciprocity, we  consider propagating SWs along the CS nanowire axis with wavenumber $k_x \neq 0$. The dispersion relations calculated for three representative values of the external magnetic field are shown in Fig.~\ref{Fig:dispersions}. The LF, HA, and HF modes identified earlier are highlighted with bold lines and colored markers. All bands shift to higher frequencies with increasing field strength, however their asymmetry with respect to $k_x =0$ varies significantly.
To trace this change, we introduce a standard measure of non-reciprocity:
\begin{equation}
    \delta{f(|k_x|)} \equiv f(-|k_x|) - f(|k_x|), 
\end{equation}
which represents the frequency difference between counter-propagating modes of the same wavelength. The dependence of $\delta{f(|k_x|)}$ is presented in Fig.~\ref{Fig:NRP_modes_diff} for the LF, HA, and HF modes under different external magnetic field strengths, in the range $|k_x| \in \langle 0,40 \rangle$~rad \textmu m$^{-1}$. The observed dependencies differ qualitatively between the three mode types. This observation, together with its physical interpretation, constitutes one of the central results of this work. 

For the LF mode [Fig.~\ref{Fig:NRP_modes_diff}(a)], $\delta{f(|k_x|)}$ reaches a maximum of approximately 0.5~GHz at $k_x \approx 30$~rad/\textmu m for $B_{\mathrm{ext}}$ = 0.5~T. With increasing field, the nonreciprocity of the LF mode decreases, eventually changes sign, and remains small, i.e., $\delta f(|k_x|)<0.25$~GHz at $k_x \leq 40$~rad/\textmu m for $B_{\mathrm{ext}}$ = 1 and 2~T. 
The different behavior of $\delta f$ at 0.5 T, compared to higher fields, correlates with the change in the LF mode profile observed in Fig.~\ref{Fig:FMR}, i.e. the transition from an edge-localised mode (Fig.~\ref{Fig:FMR}(c) and (d) at $B_\text{ext}= 1$ and 2 T, respectively) to a bulk-like mode quantized along the magnetic field direction (Fig.~\ref{Fig:FMR}b at 0.5 T).

The more striking feature appears for the HA mode [Fig.~\ref{Fig:NRP_modes_diff}(b)], where $\delta{f(|k_x|)}$ exhibits a sign change from negative to positive with increasing $|k_x|$ for all values of $B_{\mathrm{ext}}$. This reversal consistently occurs near $k_x\approx17$ rad/\textmu m. On the negative side, $\delta{f}$ reaches $-0.9$~GHz for 0.5~T, while on the positive side it approaches 0.8~GHz at higher fields. The distinct behavior of $\delta f$ in dependence on $|k_x|$ for $B_\text{ext}=0.5$~T, as compared to high fields, is also reflected in the change in the SW profile. At large magnetic fields, the maximum of the amplitude is in the central part; at 0.5~T, however, it is redistributed to the edges of the CS.  
For the HF mode [Fig.~\ref{Fig:NRP_modes_diff}(c)], $\delta f(|k_x|)$ remains negative across the entire range of $k_x$ and field strengths considered. At higher fields, the dependence is monotonic, with nonreciprocity increasing in magnitude with $k_x$. At low fields ($B_{\mathrm{ext}}$ = 0.5~T), however, the behavior becomes nonmonotonic, with the maximum nonreciprocity ($-1.9$~GHz) occurring at $k_x=20$~rad/\textmu m. 

The results reveal three key features: (a) the sign of the nonreciprocity is mode profile depended; and the sign of the nonreciprocity can change with (b) increasing wavenumber, as demonstrated for the HA mode and (c) with the change of the magnitude of the magnetic field, as shown for LF mode. We next deepen our analysis by studying how the geometry of the CS cross-section influences non-reciprocity. Since the CS is defined as the set difference of two ellipses (Fig.~\ref{Fig:structure}), systematic variation of the parameters \( a_2 \) and \( b_1 \) allows us to probe the influence of the curvature of the lower and upper surfaces of the nanowire on the nonreciprocal properties of the SW modes.

\subsubsection{Influence of the CS bottom-surface curvature\label{Sec:geometry_a}}

\begin{figure}[htb]
\centering
\includegraphics[width=\columnwidth]{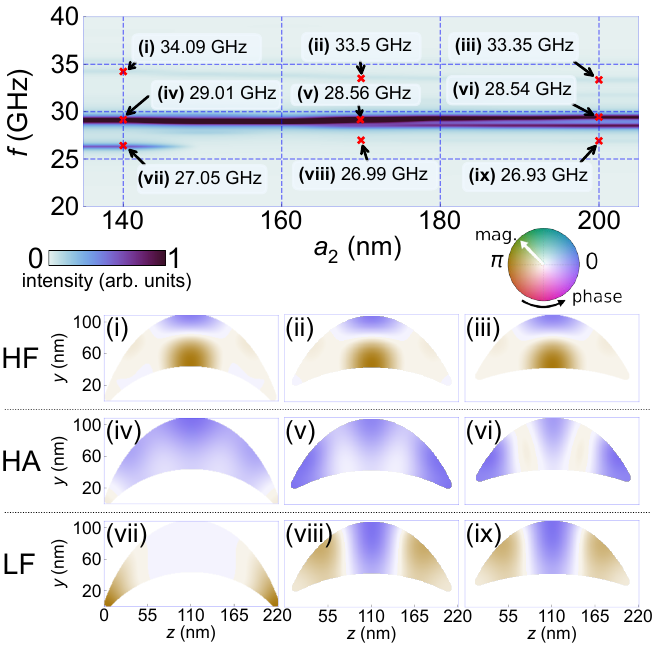}
    \caption{\label{Fig:FMR_a2}
    FMR spectra as a function of the CS semi-axis $a_2$, changed from $a_2 = 139$ to $199$~nm with step $10$~nm (see Fig.~\ref{Fig:geometry_a}), with the frequencies of the LF, HA, and HF modes indicated by red dots. The corresponding mode profiles are shown in the bottom panels, where hue encodes the local precession phase (color wheel), and opacity encodes the normalized amplitude $|m|$. The bias field $B_\text{ext}=1$~T and the profiles are shown for $k_x=0$.}
\end{figure}

We vary the parameter \( a_2 \) in the range 139--199~nm, in steps of 10~nm, while keeping the other ellipse dimensions fixed (see Fig.~\ref{Fig:geometry_a} in Appendix \ref{app:a_2}, where  the set of geometries  representing the different values of the semi-axis \( a_2 \) are shown) and under constant external magnetic field, $B_{\mathrm{ext}}$ = 1~T. This transformation progressively converts the CS nanowire into one with the bottom side of the nanowire flat. For each value of $a_2$, the FMR spectra and SW profiles were calculated, as shown in Fig.~\ref{Fig:FMR_a2}.

The LF mode become less 
intensive with increasing $a_2$ 
(see position (vii)-(ix) in Fig.~\ref{Fig:FMR_a2}). 
This decrease in the intensity is accompanied by changes in the SW profile. At $a_2=139$~nm (mode (vii)), the SW amplitude is concentrated at the edges. At $a_2=169$~nm (mode (viii)) and 199 nm (mode (ix)), the character of the mode changes and the LF mode is not an edge mode, but a bulk mode with maximal amplitude in the center and  anti-phase oscillations at the sides of the CS. In fact, similar transformation was already observed with the field decreasing in Fig.~\ref{Fig:FMR}. In both cases, increasing $a_2$ and decreasing $B_\text{ext}$ reduces the demagnetizing field at the CS edges, eliminating conditions favorable to the formation of edge modes. However, the observed effect in the case of a decreasing magnetic field is due to the rotation of the magnetization from the $z$ axis, whereas the effect in the case of an $a_2$ change is due to geometrical changes.

 As $a_2$ increases, instead of a single mode of HA at $a_2=139$ nm ($f=29.01$~GHz) there appears two at large values of $a_2$. The peak at the lower frequency is less intense and quantized along the $z$-axis.
 For further investigation  we select the one with higher frequency and stronger intensity, which amplitude distribution follows the HA mode considered so far, see the mode profiles (iv) to (v) and (vi) in Fig.~\ref{Fig:FMR_a2}.
However, we observe the change in the amplitude distribution at the cross-section. While the maximum of the HA mode in (iv) ($a_2=139$ nm) is in the center of the CS, for large values of $a_2$, e.g., (v) and (vi), the high amplitude develops at the edges and some antiphase become pronounce at $a_2=199$ nm (mode (vi)). These changes in the HA mode profile with increasing $a_2$ are again similar to the changes observed with decreasing bias magnetic field (Fig.~\ref{Fig:FMR}).
For the HF modes the intensity and the mode pattern (see modes (i)-(iii) in Fig.~\ref{Fig:FMR_a2}) remains nearly unchanged.

\begin{figure}[ht]
\centering
\includegraphics[width=\columnwidth]{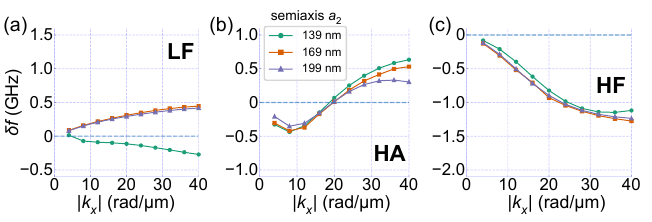}
    \caption{Nonreciprocity, $\delta f(|k_x|)=f(-|k_x|)-f(|k_x|)$, for the LF (a), HA (b), and HF (c), SW modes in three different CS geometries, i.e. for $a_2 = 139, 169$ and 199~nm. The bias field is $B_\text{ext}=1$~T.}
\label{Fig:NRP_modes_diff_a}
\end{figure}

The function  $\delta f(|k_x|)$ for three values of $a_2$ and the considered SW modes are shown in Fig.~\ref{Fig:NRP_modes_diff_a}. For the HF mode (Fig.~\ref{Fig:NRP_modes_diff_a}(c)), flattening of the bottom surface of the CS by increasing $a_2$ does not change $\delta f(|k_x|)$ dependence, as well as it doesn't alter the HF mode profile (see Fig.~\ref{Fig:FMR_a2}). For the HA mode, a change in nonreciprocity is meaningful only for larger $|k_x|$ and larger $a_2$, e.g., a decrease of $\delta f$ by ~0.25~GHz with increasing $a_2$ from 169 to 199 nm at $k_x=40$~rad/$\mu$m. This follows the change in the HA profile discussed in the previous paragraph (Fig.~\ref{Fig:FMR_a2}, mode (vi)).
The LF mode undergoes a reversal of the nonreciprocity sign for CS. It is positive for $a_2=199$ and 169 nm and negative for $a_2=139$ nm (Fig.~\ref{Fig:NRP_modes_diff_a}a). This changes are accompanied by a transition from edge localization to a bulk mode quantized along the $z$-axis (see, Fig.~\ref{Fig:FMR_a2}, modes (vii)-(ix)). 
These results demonstrate that changes in $\delta f(|k_x|)$ in dependence on $a_2$ are directly related to changes in the mode profiles discussed in the previous paragraph. 
It is worth noting that similar changes in $\delta f$ were observed with increasing $a_2$ and when $B_\text{ext}$ was decreased from 1.0~T to 0.5~T (see Fig.~\ref{Fig:NRP_modes_diff}). 

\subsubsection{Influence of the CS upper-surface curvature \label{Sec:geometry_b}}

\begin{figure}[b]
\centering
\includegraphics[width=\columnwidth]{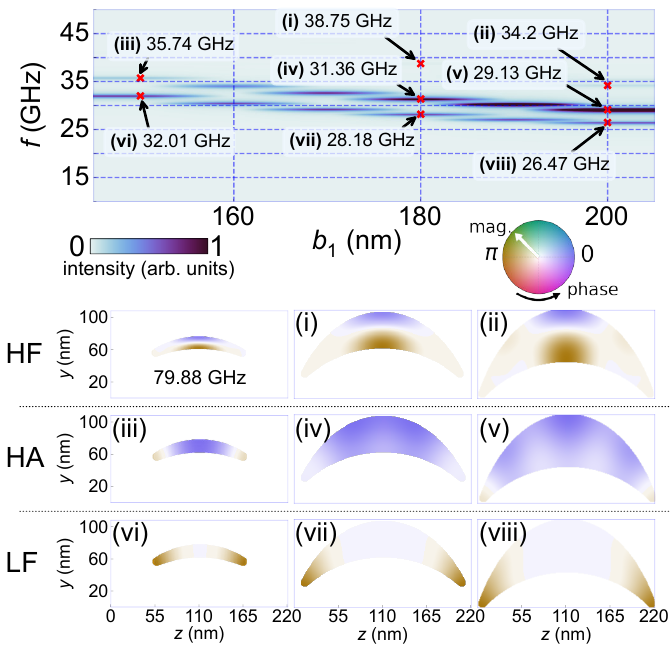}
\caption{FMR spectra as functions of the CS semi-axis $b_1 = 150$, $180$, and $200$~nm [see Fig.~\ref{Fig:geometry_b}], with the frequencies of the LF, HA, and HF modes indicated. The corresponding mode profiles are shown in the bottom panels: hue encodes the local precession phase 
(color wheel), and opacity encodes the normalized amplitude $|m|$. The bias field $B_\text{ext}=1$~T and the profiles are shown for $k_x=0$.}
\label{Fig:FMR_semiax_b}
\end{figure} 

\begin{figure}[ht]
\centering
\includegraphics[width=\columnwidth]{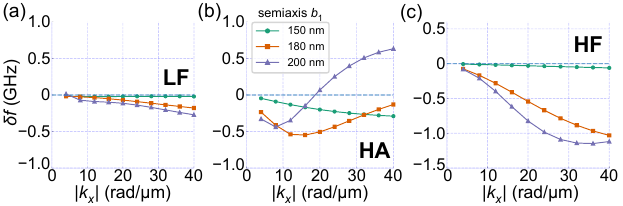}
\caption{Nonreciprocity $\delta f(|k_x|)=f(-|k_x|)-f(|k_x|)$ as functions of the wavenumber for three SW modes: LF (a), HA (b), and HF (c), calculated for three different CS geometries with $b_1 = 150, 180$ and 200~nm. The bias field $B_\text{ext}=1$~T.}
\label{Fig:NRP_modes_diff_b}
\end{figure}

We next vary \( b_1 \) in the range 200--150~nm, in steps of 10~nm and keeping fixed $B_{\mathrm{ext}}$ = 1~T (see, Fig.~\ref{Fig:geometry_b} in Appendix \ref{app:b_1} for the structure visualization). As such, we reduced maximum ellipse curvature from 0.0128~nm\(^{-1}\) to 0.0096~nm\(^{-1}\). The corresponding FMR spectra and mode profiles are presented in Fig.~\ref{Fig:FMR_semiax_b}. As $b_1$ decreases, which correspondes to a decrease in the CS thickness, all SW modes shift to higher frequencies. 
For the LF and HA modes the frequency values increase almost linearly with decreasing $b_1$, and their intensities do not vary significantly.
It is not a case of the HF mode, which is quantized along the $y$-axis. It exhibits a strong frequency increase  and  intensity decrease with decreasing $b_1$. 
The sharp increase in frequency, especially below 180 nm, is due to the dominance of exchange interactions for the modes quantized  along the $y$ axis. The loss of intensity occurs because a decrease in $b_1$ makes the cross-section of the CS more rectangular, causing the HA mode to become more antisymmetrical (see (v)-(iv)-(iii) mode profiles in Fig.~\ref{Fig:FMR_semiax_b}). Consequently, the mode's susceptibility to microwave excitation is negligible for small $b_1$.

The evolution of the SW profiles with decreasing \( b_1 \) is consistent with the progressive flattening of the CS cross-section, which approaches  more symmetric shape (see Fig.~\ref{Fig:FMR_semiax_b}).
In other words, it approaches an inversion-symmetric shape. This,  
in turn, reduces the nonreciprocity of all modes (including LF, HA, and HF modes) with decreasing $b_1$, as evidenced by the decrease of \( \delta f \) in Fig.~\ref{Fig:NRP_modes_diff_b}. However, for the HA mode (Fig.~\ref{Fig:NRP_modes_diff_b}(b)), the change in $\delta f$ with decreasing $b_1$ is not monotonic. Moreover, the dependence $\delta f(|k_x|)$ changes from non-monotonous to monotonic, to become similar to that of LF and HF (Fig.~\ref{Fig:NRP_modes_diff_b}a and c) for the smallest values of $b_1$ considered, i.e., 150~nm (green line in Fig.~\ref{Fig:NRP_modes_diff_b}b). 
We attribute these untypical changes in the course of the $\delta f(|k_x|)$ function of the HA mode to the decreasing CS thickness and the increasing frequency separation between the modes quantized along the $y$-axis. This is particularly due to the elimination of hybridization between the modes within the considered frequency range.
 This, indicates that the nonmonotonic dependence and sign change of nonreciprocity on $|k_x|$ can be due to the hybridisation of the HA mode with thickness-quantized modes. We will return to this hypothesis in the discussion section.

Our analysis shows that non-reciprocity in CS nanowire is not a uniform property, but rather, it depends on the mode character and the interaction between the modes. This character can change due to variations in geometry and the    bias magnetic field magnitude, but may also change with wavenumber. To shed more light on how nonreciprocity depends on the mode character, we propose a phenomenological model.

\subsection{Ellipticity-driven sign reversal of nonreciprocity}
\label{Sec:EllipticityMain}
Let us assume that the external field \(H\) (\( \mu_{0} H = B_{\mathrm{ext}} \)) is applied along the $z$ axis, and it is strong enough to consider the static magnetization parallel to it in whole volume of the nanowire. Taking into account the strong spatial inhomogeneity of the demagnetizing fields~\cite{CRPHYS_2005,/doi.org/10.1002/}, this can only be a working approximation. 
However, deviations of the magnetization from the \(z\) direction make a quantitative, but not qualitative, effect on the analysis of possible nonreciprocity that we carry out here.

In the SW approximation, we assume that the magnetization vector has the following form
\begin{equation}
\mathbf{M}=\Big(A\,m_x(y,z)\,e^{-i\Omega t + ik_xx},\;
               B\,m_y(y,z)\,e^{-i\Omega t + ik_xx},\;
               M_\text{s}\Big),
\label{eq:M_ansatz}
\end{equation}
where \(\Omega = 2\pi f\) is angular frequency, the SW profiles in the transverse cross-section of the nanowire, i.e., in the \((y,z)\) plane, are: \(m_x(y,z)\) and \(m_y(y,z)\). The frequency and the SW profile
depend on the wave vector \(k_x\), but to simplify the notation, we will drop the $k_x$ index in the following unless it is necessary to keep it.

The corresponding Landau–Lifshitz (LL) equations are
\begin{equation}
\frac{i\Omega}{\gamma\mu_0}\,A\,m_x
= B\big[-M_\text{s} D\,\Delta_{}+H\big]\,m_y - M_\text{s}\,h_y, 
\label{eq:LLx}
\end{equation}
\begin{equation}
-\frac{i\Omega}{\gamma\mu_0}\,B\,m_y
= A\big[-M_\text{s}D\,\Delta_{}+H\big]\,m_x - M_\text{s}\,h_x ,
\label{eq:LLy}
\end{equation}
where \(\Delta_{}=\partial_y^2+\partial_z^2-k_x^2\) denotes the Laplacian operator, \(\sqrt{D} = \sqrt{{2A_{\mathrm{ex}}}/{\mu_{0} M_{s}^2}}\) is the exchange length.
The $x$ and $y$ components of the
dynamic dipolar field, $h_x(y,z)$ and $h_y(y,z)$, take the form
\begin{align}
h_x(y,z) &= -M_\text{s} \,e^{ik_xx}\!\!\iint dy'\,dz'\,
\Big( k_x^2 A\,m_x(y',z')\,\Psi_{k_x} \nonumber \\
&\quad +\, i k_x B\,m_y(y',z')\,\partial_{y'} \Psi_{k_x} \Big), 
\label{eq:hx_app}\\[4pt]
h_y(y,z) &= -M_\text{s}\,e^{ik_xx}\!\!\iint dy'\,dz'\,
\Big(- i k_x A\,m_x(y',z')\,\partial_{y'} \Psi_{k_x} \nonumber \\
&\quad +\, B\,m_y(y',z')\,\partial_{y'}^{2} \Psi_{k_x} \Big),
\label{eq:hy_app}
\end{align}
where the kernel $\Psi_{k_x}$ is an even function of $k_x$, and is defined as
\begin{equation}
\Psi_{k_x}(y-y',z-z')=\int_{0}^{\infty}
\frac{2\cos(k_x t)\,dt}{\sqrt{t^2+(y-y')^2+(z-z')^2}} .
\label{eq:Psi_app}
\end{equation}

Inserting $h_x$ and $h_y$ into Eqs.~\eqref{eq:LLx}–\eqref{eq:LLy} yields two
integro–differential LL equations for $m_x(y,z)$ and $m_y(y,z)$. An exact analytic solution
is generally not possible in the presence of strong dipolar inhomogeneities~\cite{vazquez2020magnetic}.
Instead, we employ a variational approach similar to the Ritz ansatz, seeking trial
functions guided by numerical simulations~\cite{PhysRevB.96.024446}.
Averaging over the cross-section leads to
\begin{align}
\frac{i \Omega}{\gamma \mu_0} A \langle m_x^* m_x \rangle &= 
B \left[ \langle m_x^* \hat{H} m_y \rangle - h_{yy}\right] - i k_x A h_{yx}, \label{eq:avg1} \\
- \frac{i \Omega}{\gamma \mu_0} B \langle m_y^* m_y \rangle &= 
A \left[ \langle m_y^* \hat{H} m_x \rangle - k_x^2 h_{xx}\right] - i k_x B h_{xy}, \label{eq:avg2}
\end{align}
where star in \(m_i\)  indicates the complex conjugate and 
$h_{ij}$ is a dynamical dipolar tensor (see Eq.~\eqref{eq:hii} in Appendix). We introduce the notation
\begin{equation}
\langle F\rangle=\frac{1}{S}\int\!\!\int F(y,z)\,dy\,dz,
\label{eq:average}
\end{equation}
the integration is over the cross-sectional area, $S$. The operator $\hat{H}$ acting on the trial functions is
\begin{equation}
\hat{H}=-M_\text{s}D\Delta_{}+H.
\label{eq:Hhat}
\end{equation}
Setting the determinant of Eqs.~\eqref{eq:avg1}–\eqref{eq:avg2} to zero gives
\begin{align}
\begin{aligned}
&\left(\frac{\Omega}{\gamma \mu_0}\right)^{2} 
\langle m_x^{*} m_x \rangle \langle m_y^{*} m_y \rangle \\ 
&- \frac{\Omega}{\gamma \mu_0} k_x 
\Big( \langle m_x^{*} m_x \rangle h_{xy} - \langle m_y^{*} m_y \rangle h_{yx} - k_x^2 h_{yx} h_{xy} \Big) \\
&- \Big( \langle m_x^{*} \hat{H} m_{y} \rangle - h_{yy} \Big) 
   \Big( \langle m_y^{*} \hat{H} m_{x} \rangle - k_x^2 h_{xx} \Big) = 0 .
\end{aligned}
\label{eq:disp}
\end{align}
To analyze the origin of the nonreciprocity, we examine Eq.~\eqref{eq:disp}, from which the relevant mechanisms directly follow: 
(i) A secular term linear in $k_x$ and $\Omega$, absent in axially symmetric systems,
appears here due to static-magnetization asymmetry~\cite{Ishibashi2020,  10.1063/10.0038642}. 
(ii) Nonreciprocity also arises from different SW profiles for $\pm k_x$, due to inhomogeneous demagnetizing fields~\cite{CRPHYS_2005, /doi.org/10.1002/}.

Large $B_{\mathrm{ext}}$ minimizes static dipolar inhomogeneity, make profiles nearly uniform compared to non-local factor  $\Psi_{k_x}(y-y',z-z')$, Eq.~\eqref{eq:Psi_app}, in the central part of the cross-section (for instance, for the LF mode). In this case the term proportional to $\langle m_x^{*} m_y \rangle h_{xy} - \langle m_y^{*} m_y \rangle h_{yx}$ is almost equal to zero, and, using Eq.~\eqref{eq:disp}, we can write for nonreciprocity
\begin{equation}
   \frac{1}{\gamma\,\mu_{0}} \left(\Omega_{-|k_x|}- \Omega_{|k_x|}\right)
    = \bigl(\gamma_{-|k_x|}-\gamma_{|k_x|}\bigr)\,F\!\left(k_x^{2}\right),
    \label{eq:nonrec}
\end{equation}
where $F\!\left(k_x^{2}\right)$ is an even function of $k_x$ (see, Appendix B, Eq.~\eqref{eq:freq_k}) and
the modal ellipticity factor $\gamma_{k_x}$ is given by:
\begin{equation}
    \gamma_{k_x} \;=\;
    \frac{\bigl\langle m_x^*(y,z)\, m_y(y,z) \bigr\rangle\,
          \bigl\langle m_y^*(y,z)\, m_x(y,z) \bigr\rangle}
         {\bigl\langle |m_x(y,z)|^2 \bigr\rangle\,
          \bigl\langle |m_y(y,z)|^2 \bigr\rangle}.
    \label{eq:gamma_def}
\end{equation}

\begin{figure}
\centering
\includegraphics[width=\columnwidth]{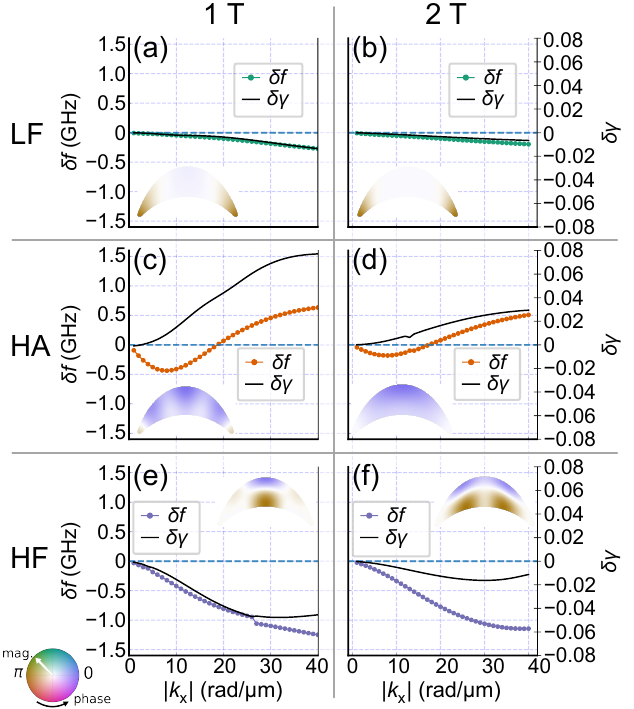}
    \caption{Comparison of the nonreciprocal frequency shift $\delta f(|k_x|)$ and the modal ellipticity asymmetry $\delta\gamma(|k_x|)$ for: (a-b) the LF, (c--d) HA, and (e--f) HF  modes. The $\delta f(|k_x|)$ and $\delta\gamma(|k_x|)$ are obtained from simulations and with the use of Eq.~\ref{eq:gamma_def}, respectively, at two values of the external magnetic field 1~T (a,c,e) and 2~T (b,d,f).  Insets illustrate representative mode profiles for $k_x=0$, hue encodes the local precession phase (see a color wheel), and opacity encodes the normalized amplitude $|m|$.}
\label{Fig:gamma_plot}
\end{figure}

We numerically evaluate the quantity $\delta\gamma(|k_x|) \equiv \gamma_{-|k_x|} - \gamma_{|k_x|}$ as a proxy for nonreciprocity, $\delta f(|k_x|)$. Fig.~\ref{Fig:gamma_plot} shows a comparison of the two functions for three types of SW modes at two bias magnetic fields: 1 T and 2 T. For the LF branch [Fig.~\ref{Fig:gamma_plot}(a,b)], $\delta\gamma(k_x)$ closely follows $\delta f(|k_x|)$ across the entire $k_x$-range. For the HA mode the $\gamma(|k_x|)$ dependence also reflects general nonreciprocity dependence but without a sign change, i.e., the point where $\delta f(|k_x|)$ crosses zero (near $|k_x|\!\approx\!17$~rad/\textmu m)  [Fig.~\ref{Fig:gamma_plot}(c,d)]. For the HF mode, [Fig.~\ref{Fig:gamma_plot}(e,f)], both $\delta \gamma$ and $\delta f$ show a similar behavior on wavenumber, with the same sign of the nonreciprocity.
 
\begin{figure}
\centering
\includegraphics[width=\columnwidth]{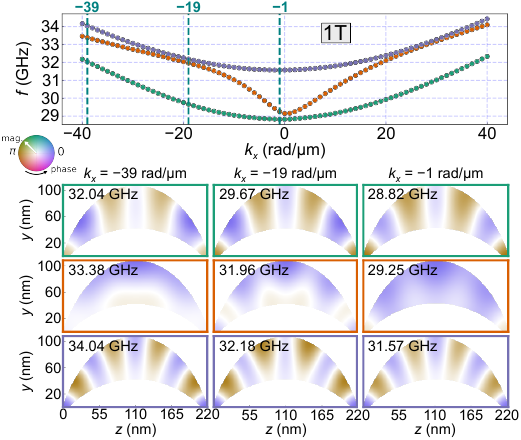}
\caption{ The dispersion relation of HA SW (red dots) and two other modes in the CS nanowire ($a_1=125$~nm, $b_1=200$~nm, $a_2=139$~nm and $b_2=135$~nm) in magnetic field of 1 T. The bottom panels show spatial distributions of the dynamic magnetization amplitude $|m|$ at $k_x = -1.0$, $-19.0$, and $-39.0$~rad/\textmu m. Hue encodes the local precession phase (color wheel), while opacity indicates the normalized amplitude (transparent at 0, opaque at 1). }
\label{Fig:hybrid1000}
\end{figure}

\subsection{Discussion}
Intriguing observation of this study is the change in the sign of nonreciprocity for the HA mode with increasing $k_x$, visible in simulation results in Figs.~\ref{Fig:NRP_modes_diff}(b), \ref{Fig:NRP_modes_diff_a}(b) and \ref{Fig:NRP_modes_diff_b}(b) (only for $b_1=200$ nm), but not observed in the $\delta \gamma(|k_x|)$ dependence  (Fig.~\ref{Fig:gamma_plot} (c) and (d)).
A change in the sign of the $\delta f$ with respect to the wavenumber was previously observed in a thin film with a gradient in the static magnetization value along the thickness for SWs propagating perpendicular to the magnetization direction, i.e., in the Damon-Eshbach configuration \cite{Gallardo_2019}. In fact, two such critical points have been identified: one at low wavenumbers (around $k=10$ rad/\textmu m), and the other at high wavenumbers (around $k=160$ rad/\textmu m). The appearance of the first critical point was correlated with the quantized mode along the film thickness, perpendicular standing SW (PSSW). As the film thickness increases (from 30 \textmu m), the critical point shifts to lower wavenumbers and eventually disappears. Conversely, the second critical point shifts to higher wavenumbers with increasing film thickness and it was correlated with the transition from a dipole-dominating regime to an exchange-dominating regime.
Similar asymmetric hybridization between the PSSW and the fundamental modes was observed in the homogeneous ferromagnetic film but with asymmetric surface anisotropies~\cite{Szulc2024}. While $\delta f$ was not the focus of this study, the results clearly indicate a direct corelation between the hybridization of the fundamental mode and the PSSW on nonreciprocity. Recently, the strong asymmetric hybridization between the fundamental magnetostatic mode and first-order PSSW was demonstrated in bilayers composed of two ferromagnetic films in direct contact~\cite{PhysRevB.111.134434}. The change of the sign of the nonreciprocity has also been shown experimentally in the dipole-coupled two \ce{FeGa} films when the relative structural asymmetry was changed by tuning the sublayer thicknesses at a fixed wavenumber ~\cite{Gerevenkov2023}. At the critical point, the sign of $\delta f$ for acoustic and optical mode exchanges.

The situation is more complex in the case of the CS nanowire, studied in this paper. The sign of the nonreciprocity parameter, $\delta f$, is mode-dependent. It is positive for the LF mode, when $B_\text{ext}=0.5$~T (Fig.~\ref{Fig:NRP_modes_diff}a) or at higher field values, but only when $a_2$ is increased above 170 nm (Fig.~\ref{Fig:NRP_modes_diff_a}a).  This behavior implies that $\delta f > 0$ for the LF mode when it has a bulk-type character, which exhibits quantization along the $x$-axis. In contrast, for the HF mode  which is also a bulk mode but quantized along the $y$-axis, or for the LF mode when it is an edge-localized mode,  $\delta f < 0$.
The situation is different for the fundamental mode, i.e., the HA mode, where the sign of the $\delta f$ changes with $k_x$ for the most studied case (except for the CS with $b_1 \leq 180$ nm, when the CS cross-section transforms towards a thin rectangle shape. Fig.~\ref{Fig:NRP_modes_diff_b}b). As previously mentioned, this suggests that the critical point (where $\delta f=0$) at $k_x \neq 0$ is associated with the hybridization of the HA with the other SW modes in the system. This is further supported by the comparison of the functions $\delta f(|k_x|)$ and $\delta \gamma(|k_x|)$ in Fig.~\ref{Fig:gamma_plot}c and d, which show that the phenomenological model in the SW approximation (where interactions between SW modes are excluded) makes no predictions of a critical point.

To further support this hypothesis we  show in Fig.~\ref{Fig:hybrid1000} the dispersion relation of the HA mode (red dots) and the two closest SW modes in the CS nanowire (of starting dimensions) at $B_\text{ext}=1.0$~T. There is a clear variation in the HA mode dispersion for negative wavenumbers, where the HA mode band almost touches the higher mode dispersion at $k_x=-20$~rad/\textmu m (which is close to the critical point, $|k_x|=19$~rad/\textmu m, where $\delta f (|k_x|)=0$, see, Fig.~\ref{Fig:NRP_modes_diff}b), but does not cross it. The profile of the HA mode also changes, with anti-phase oscillations appearing in the cross-sections along the $z$ and $y$ axis at negative wavenumbers. This indicates on the multi-mode hybridization between HA and higher order SW modes. 


\section{Conclusions}

We numerically studied the SW propagation in a long CS nanowire magnetized perpendicular to its axis, focusing on the asymmetry of the dispersion relation. Our results show that the type of SW is an important factor influencing nonreciprocity, which is measured as the difference in frequencies of waves with the same wavenumber that propagate in opposite directions. Specifically, the sign of the nonreciprocity (under the assumed convention) is negative for the edge mode and the mode quantized along the CS width, but positive for the mode quantized along the bias field. Interestingly, changing the CS geometry (i.e., changing the curvature of the bottom or top CS surface as well the magnitude of the bias magnetic field) can induce a change in the sign of the nonreciprocity for the lowest frequency mode. This transition is associated with a change in mode type (e.g., from an edge mode to a bulk mode) due to a change in the demagnetizing field at the CS edges. This change can be caused by a change in geometry or rotation of the magnetization.
Furthermore, we observed a change in the sign of the nonreciprocity in dependence on the wavenumber for the fundamental mode. As with the previously studied asymmetric thin films, we demonstrate that this effect originates from the hybridization of the fundamental SW mode with higher-frequency, thickness-quantized SW modes. In the case of CS, however, it is multimode hybridization.
These results demonstrate the interesting tunability of nonreciprocal effects in CS nanowires, which depends on the SW mode type. The results highlight the potential of nanowires with a CS cross-section as building blocks for magnonic devices, especially that they can be fabricated with current technologies and the three main types of SW modes can be measured with standard broadband FMR techniques.

\begin{acknowledgments}
The research leading to these results was funded by the National Science Centre of Poland, Projects: OPUS Nos.~UMO-2023/49/B/ST3/02920 and UMO-2020/39/I/ST3/02413, PRELUDIUM Nos.~UMO-2024/53/N/ST3/03244 and~UMO-2023/49/N/ST3/03032. The contribution of the Norwegian Financial Mechanism 2014-2021, Project POLS No. UMO-2020/37/K/ST3/02450 is also acknowledged.
\end{acknowledgments}
\section*{Data Availability Statement}
The data that support the findings of this study are openly available on Zenodo at \url{https://doi.org/10.5281/zenodo.17296216}.
\appendix
\section{Geometry}
\label{app:geometry}
\subsection{Varying of the CS bottom surface}\label{app:a_2}
\begin{figure}[htbp!]
  \centering
\includegraphics[width=0.42\textwidth]{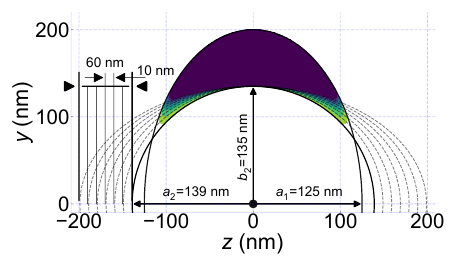}
  \caption{Geometry of CS nanowires with varying semi-axis $a_2$ (139--199 nm). }
  \label{Fig:geometry_a}
\end{figure}

Figure~\ref{Fig:geometry_a} illustrates the CS geometry with varying the semi-axis $a_2$ (139--199 nm), which modifies the bottom curvature of the CS nanorod. 

\subsection{Varying of the CS top surface}\label{app:b_1}
\begin{figure}[htbp!]
  \centering
\includegraphics[width=0.42\textwidth]{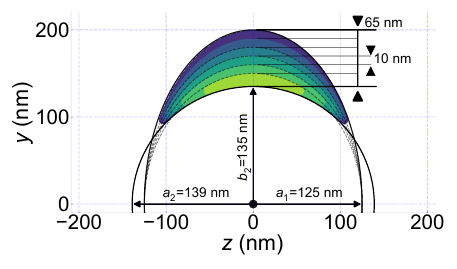}
  \caption{Geometry of CS nanowires with varying semi-axis $b_1$ in the range from 150 to 200 nm. On the right the thicknesses of the resulting CS are indicated.}
  \label{Fig:geometry_b}
\end{figure}

Figure~\ref{Fig:geometry_b} illustrates the CS geometries with varying the outer semi-axis $b_1$ in the range 150--200 nm. It modifies the top curvature and the thickness of the nanowire. 

\section{Dynamic dipolar field elements}
\label{app:gamma}

\begingroup
\setlength{\abovedisplayskip}{4pt}
\setlength{\belowdisplayskip}{4pt}
\setlength{\abovedisplayshortskip}{3pt}
\setlength{\belowdisplayshortskip}{3pt}
\setlength{\jot}{2pt}

Below we provide the explicit formula on the dipole tensor elements
$h_{ij}$ that enter Eqs.~\eqref{eq:avg1}–\eqref{eq:avg2} 
in the main text:
\begin{align}
\begin{aligned}
h_{yy}&=-M_\text{s}\,\Big\langle m_x^*\!\!\iint dy'\,dz'\;
m_y(y',z')\,\partial_{y'}^{2}\Psi_{k_x} \Big\rangle, \\
h_{yx}&= M_\text{s}\,\Big\langle m_x^*\!\!\iint dy'\,dz'\;
m_x(y',z')\,\partial_{y'}\Psi_{k_x} \Big\rangle, \\
h_{xx}&=-M_\text{s}\,\Big\langle m_y^*\!\!\iint dy'\,dz'\;
m_x(y',z')\,\Psi_{k_x} \Big\rangle, \\
h_{xy}&= M_\text{s}\,\Big\langle m_y^*\!\!\iint dy'\,dz'\;
m_y(y',z')\,\partial_{y'}\Psi_{k_x} \Big\rangle.
\end{aligned}
\label{eq:hii}
\end{align}
In the case of nearly uniform profile of SW in the $zy$ plane,
\begin{align}
\begin{aligned}
h_{yy} &\approx -M_\text{s} \, \langle m_x^{*} m_y \rangle 
\left\langle \iint dy' \, dz' \, \frac{\partial}{\partial y'} \frac{\partial}{\partial y} 
\Psi(y, z, y', z') \right\rangle, \\
h_{xx} &\approx -M_\text{s} \, \langle m_y^{*} m_x \rangle 
\left\langle \iint dy' \, dz' \, \Psi(y, z, y', z') \right\rangle, \\
h_{yx} &\approx M_\text{s} \, \langle m_x^{*} m_x \rangle 
\left\langle \iint dy' \, dz' \, \frac{\partial}{\partial y} 
\Psi(y, z, y', z') \right\rangle, \\
h_{xy} &\approx M_\text{s} \, \langle m_y^{*} m_y \rangle 
\left\langle \iint dy' \, dz' \, \frac{\partial}{\partial y} 
\Psi(y, z, y', z') \right\rangle.
\end{aligned}
\label{eq:hii_approx}
\end{align}
Also, we can approximate:
\begin{align}
\begin{aligned}
&\langle m_x^{*} m_x \rangle h_{xy} - \langle m_y^{*} m_y \rangle h_{yx} \approx 0, \\[6pt]
&\langle m_x(y,z)^{*} \, \tilde{H} m_{y}(y,z) \rangle \approx H_1\,\langle m_x^{*} m_y \rangle, \\[6pt]
&\langle m_y(y,z)^{*} \, \tilde{H} m_{x}(y,z) \rangle \approx H_1\,\langle m_y^{*} m_x \rangle .
\end{aligned}
\label{eq:hm-rel}
\end{align}
where \((H +M D k_x^2)\) is defined as \(H_1\).

In such a case using~\eqref{eq:hii_approx} and approximations \ref{eq:hm-rel}, we can rewrite the equation~\eqref{eq:disp} as
\begin{equation}
\begin{aligned}
\left( \frac{\Omega}{\gamma \mu_{0}} \right)^{2} 
= k_{x}^{2} M_\text{s}^{2} 
\left( \left\langle \iint dy' \, dz' \, 
\frac{\partial}{\partial y} \Psi(y,z;y',z') \right\rangle \right)^{2} \\[6pt]
+ \frac{\bigl\langle m_x^*(y,z)\, m_y(y,z) \bigr\rangle\,
          \bigl\langle m_y^*(y,z)\, m_x(y,z) \bigr\rangle}
         {\bigl\langle m_x^*(y,z)\, m_x(y,z) \bigr\rangle\,
          \bigl\langle m_y^*(y,z)\, m_y(y,z) \bigr\rangle} \\[6pt]
\times
\Bigg[ 
\left( H_1 
+ M_\text{s} \left\langle \iint dy' dz' \, \frac{\partial}{\partial y'} \frac{\partial}{\partial y} \Psi(y,z;y',z') \right\rangle \right) \\
\quad \times 
\left( H_1 
+ k_{x}^{2} M_\text{s} \left\langle \iint dy' dz' \, \Psi(y,z;y',z') \right\rangle \right) 
\Bigg]
\end{aligned}
\label{eq:disp_approx}
\end{equation}

By writing the Eq.~\eqref{eq:disp_approx} for $\Omega_{|k_x|}$ and $\Omega_{-|k_x|}$, we obtain expression~\eqref{eq:nonrec} for the nonreciprocity and the following for the even function $F(k_x^2)$:
\begin{align}
F(k_x^2) =
{
\Big(H_1 + M_\text{s} \left\langle \iint dy' \, dz' \, 
\frac{\partial}{\partial y'} \frac{\partial}{\partial y} 
\Psi(y,z,y',z') \right\rangle \Big)}  \notag \\
\times
\frac{\Big( H_1 + k_x^2 M_\text{s} \left\langle \iint dy' \, dz' \, 
\Psi(y,z,y',z') \right\rangle \Big)
}{
\dfrac{\Omega_{|k_x|}}{\gamma H_0} + \dfrac{\Omega_{-|k_x|}}{\gamma H_0}
}
\label{eq:freq_k}
\end{align}
\endgroup

\bibliography{refs}
\end{document}